# Strain in epitaxial high-index Bi2Se3(221) films grown by Molecular-Beam Epitaxy

B. Li[1], W. G. Chen[2], X. Guo[1], W. K. Ho[1], X. Q. Dai[2], J. F. Jia[3], and M. H. Xie[1,a)]

[1]*Physics Department, The University of Hong Kong, Pokfulam Road, Hong Kong*

[2]*College of Physics and Electronic Engineering, Henan Normal University, Xinxiang, Henan 453007, China; and School of Physics and Electronic Engineering, Zhengzhou Normal University, Zhengzhou, Henan 450044, China*

[3] *Key Laboratory of Artificial Structures and Quantum Control (Ministry of Education); Collaborative Innovation Center of Advanced Microstructures; Department of Physics and Astronomy, Shanghai Jiaotong University, 800 Dongchuan Road, Shanghai, 200240, China.*

Abstract. High-index $Bi_2Se_3$(221) film has been grown on $In_2Se_3$-buffered GaAs(001), in which a much retarded strain relaxation dynamics is recorded. The slow strain-relaxation process of in epitaxial $Bi_2Se_3$(221) can be attributed to the layered structure of $Bi_2Se_3$ crystal, where the epifilm grown along [221] is like a pile of weakly-coupled quintuple layer slabs stacked side-by-side on substrate. Finally, we have revealed the strong chemical bonding at the interface of $Bi_2Se_3$ and $In_2Se_3$ by plotting differential charge contour calculated by first-principle method. This study points to the feasibility of achieving strained TIs for manipulating the properties of topological systems.

Topological property of matter is one of the most intriguing and important aspects of condensed matter physics that has attracted extensive research attentions in recent years.[1–12] Examples include quantum Hall and quantum spin Hall systems,[1–4] topological insulators (TIs),[5,6] Dirac and Weyl semimetals.[7,8] Searching topological systems remains one of the main themes in condensed matter research and many materials have now been identified as the TIs (e.g., $Bi_2Se_3$,[9] $Bi_2Te_3$ [10,11] and $Sb_2Te_3$ [11])[12] or Dirac/Weyl semimetals (CdAs[7] and TaAs[8]), for example. The prevailing feature of these topological materials is the existence of Dirac cone dispersion, which

---

a) Email: mhxie@hku.hk

promises massless Dirac electronics. It will be scientifically interesting to manipulate Dirac electrons via, e.g., the topological phase transition, by tuning the external constraints. One of the constraints is strain which may be applied or removed reversibly.[13] Theoretical studies have predicted that the TI phase of $Bi_2Se_3$ under the ambient condition can be transformed to a topological semimetal and an ordinal insulator (OI) phase by application of a tensile strain of ~ 7% along [111], the *c*-axis direction.[14,15] The rhombohedral phase of $Sb_2Se_3$, on the other hand, may be tuned from OI to TI when compressed by a similar magnitude.[14] Experimentally, uniformly strained $Bi_2Se_3$ and $Sb_2Se_3$ at such magnitudes can hardly be realized due to their layered crystal structures. Little work is thus available to examine strain-induced topological phase transition by experiments except for a scanning tunneling microscopy (STM) study of a local strain effect induced by defects in the TI epilayer.[16] On a lower scale, on the other hand, strain introduced by mechanically bending a thin film has been shown to tune the Fermi level effectively.[17]

Theoretical studies have suggested that the helical Dirac cone has a unique anisotropic elliptical shape on $Bi_2Se_3(221)$, which is rooted from the symmetry of its surface lattice.[18,19] Such elliptical Dirac cone has been experimentally revealed also by angle-resolved photoelectron spectroscopy measurements.[20] The group velocity ($\bar{v} = \frac{1}{\hbar}\frac{dE}{dk}$) of the Dirac electrons is found slower than that of $Bi_2Se_3(111)$. A recent first principles calculation further indicates that the helical Dirac cone of $Bi_2Se_3(221)$ will experience a prominent anisotropic deformation under strain along the in-plane a-axis, where the group velocity is tuned by different magnitudes along different crystallographic directions and the response is more sensitive to compressive strain than the tensile one (e.g., a compressive strain of 3% already brings about significant changes of the Dirac cone dispersion). This is in addition to the Fermi level tuning by strain echoing that of Ref. [17]. All of these suggest the important roles strain may play in manipulating Dirac electrons.

A common strategy to achieving a uniformly strained film is to grow it epitaxially on a substrate with adequate lattice misfit, where a thin epilayer becomes stressed by the substrate lattice for coherent growth. According to elasticity theory, a strained coherent epifilm can be favored over the dislocation-mediated strain-relieved one when the film thickness is below a critical thickness.[21] This is because creating strain-relieving misfit dislocations would invoke breaking chemical bonds at the hetero-interface, which can be more energy costly than stressing the film of thin thicknesses. However, for layer-structured crystals like $Bi_2Se_3$ and $Sb_2Se_3$, which contain weakly bonded atomic planes via van der Waals (vdW) forces (e.g., between [Se-Bi-Se-Bi-Se] quintuple layers (QLs) in $Bi_2Se_3$), strain relaxation may not necessarily invoke breaking chemical bonds but occur readily at the vdW 'gaps'. It is thus difficult to achieve strained epifilms of $Bi_2Se_3$ when grown along the *c*-axis direction.[22] It is unfortunate that growth of Bi$_2$Se$_3$ preferably proceeds along *c*-direction irrespective of the substrates used.[23,24] Therefore, strain-free Bi$_2$Se$_3$ epifilms almost always result even for very thin layers. The latter is in fact characteristic of the so-called vdW epitaxy, which allows growths of the layered compounds on substrates of large lattice misfits.[25]

In an early experiment, we demonstrated successful growth of high-index Bi$_2$Se$_3$(221) epifilm by molecular-beam epitaxy (MBE) on purposely roughened InP(001) substrate.[20] The off *c*-axis growth direction of Bi$_2$Se$_3$ occurred because of the epitaxial relation $(111)_{Bi_2Se_3}||(111)_{InP}$, where the $(111)_{InP}$ facets were generated by chemical and thermal roughening of the (001) surface of the substrate. The *c*-, or {111}, planes of Bi$_2$Se$_3$ are thus inclined with respect to the $(221)_{Bi_2Se_3}||(001)_{InP}$ surface by ~57.9° and atoms at the hetero-interface are chemically bonded [refer to Fig. 1(d)]. Such a film is like a pile of weakly coupled QL slabs stacked side-by-side on substrate, which can be more tolerable to strain even than that of a covalent crystal. In the earlier experiment using InP as the substrate,[20] strain was not of primary concern and there was little lattice misfit between the deposit and substrate in the *a*-axis direction. Here in this work, we report growth of strained Bi$_2$Se$_3$(221) on In$_2$Se$_3$-buffered GaAs(001) substrate and note a much retarded strain-relaxation dynamics. This experiment demonstrates the feasibility of obtaining strained TI films, at least for strains ≤ 3%, which would lead to effective Fermi level and group velocity tuning as suggested by first principles calculation.

Vienna Ab-initio Simulation Package (VASP) was employed to calculate the charge transfer during the formation of chemical bonds at the interface. Monkhorst-Pack method was used to generate 1×5×5 reciprocal mesh. The cut-off energy was 340eV and spin-orbit coupling is on. Perdew-Burke-Ernzerhof (PBE) potential was applied for exchange correlation.

The MBE system in which film deposition experiments were carried out had a background pressure of $10^{-10}$ torr. It was equipped with a reflection high-energy electron diffraction (RHEED) facility operated at 10 keV. The diffraction pattern formed on phosphorous screen were captured by a high-speed CCD camera (AVT Stingray F046B) with the intensity data acquired by a 1394 card and processed by a desktop computer. Commercial GaAs(001) wafers were etched in the solution of H$_2$SO$_4$+H$_2$O$_2$+deionized water before being loaded into the vacuum system followed by an annealing at ~550 °C for 2.5 hours until the RHEED pattern became bright and spotty as shown in Fig. 1(a). Such a RHEED pattern indicated a clean but roughened surface of GaAs(001). Instead of depositing Bi$_2$Se$_3$ directly on the GaAs substrate, an In$_2$Se$_3$ buffer layer was grown at 500 °C, which was to achieve better quality of Bi$_2$Se$_3$ subsequently grown on top (i.e., for reduced defect density and rotation domains). During In$_2$Se$_3$ deposition, the RHEED pattern was seen to develop gradually from spotty to diamond-like showing inclined line features connecting the diffraction spots [see Fig. 1(b)]. We refer to the latter as the 'linked-spots' pattern, which was found advantageous for achieving uniform and single domain $Bi_2Se_3$(221) films on top. After the "linked-spots" RHEED patterns was fully developed, $In_2Se_3$ deposition was interrupted while $Bi_2Se_3$ growth was initiated upon the sample temperature settled at 200 °C. For both $Bi_2Se_3$ and $In_2Se_3$ growth, the source fluxes generated from K-cells were set at the ratio of 1 : 10 between metal (In and Bi) and Se. Film growth rate was 0.26 nm/min estimated by post-growth film thickness measurements by cross-sectional transmission electron microscopy (TEM). During MBE, the RHEED pattern and the spacing '$D$' between first-order diffraction spots/streaks were recorded as function of deposition time.

We remark firstly that the buffer film was α-phase $In_2Se_3$ as inferred from the symmetry and inter-diffraction spot spacing measurements of the diffraction patterns. The α-phase $In_2Se_3$ has the similar layered rhombohedral crystal structure as $Bi_2Se_3$ but with a different lattice constant (i.e., $a_{In_2Se_3} \approx 4.025$ Å and $c_{In_2Se_3} \approx 28.76$ Å, which compares that of $Bi_2Se_3$: $a_{Bi_2Se_3} \approx 4.14$ Å and $c_{Bi_2Se_3} \approx 28.64$ Å). The lattice misfit amounts to about ∼3% along the *a*-axis. This is significantly larger than that of $Bi_2Se_3$-on-InP (∼0.2%), allowed us to follow the strain by *in situ* RHEED. We also characterized the grown samples by room-temperature (RT) STM under a constant tunneling current of $I_t = 0.1$ nA and sample bias of $V_s = 1.0$ V. Cross-sectional high-resolution TEM and selective area electron diffraction (SAED) measurements were also performed using a FEI Tecnai G2 20 S-TWIN scanning transmission electron microscope. The TEM specimens were prepared following the standard procedure of mechanical thinning and argon ion milling.

Fig. 1(c) shows the RHEED pattern taken from a 22nm-thick $Bi_2Se_3$ film grown on $In_2Se_3$-buffered GaAs(001) substrate. It is seen that the "linked-spots" pattern of the starting surface has changed to one of superimposing spots and vertical streaks and the inclined line features of the starting surface have faded out. The spots show the symmetry of $Bi_2Se_3$ viewed along $[11\bar{4}]$ and measurements of inter-streak spacing suggest an in-plane lattice constant consistent with $Bi_2Se_3$. Unexpectedly, we also observe fractional streaks [labeled by "I" in Fig. 1(c)] that are persistent during the whole deposition process. Such fractional diffraction streaks imply the existence of a surface superstructure with a periodicity double that of $Bi_2Se_3$ bulk lattice. We do not yet know the nature and origin of such superstructure on epitaxial $Bi_2Se_3$ but merely remark that it is quite reproducible and does not appear an artefact.

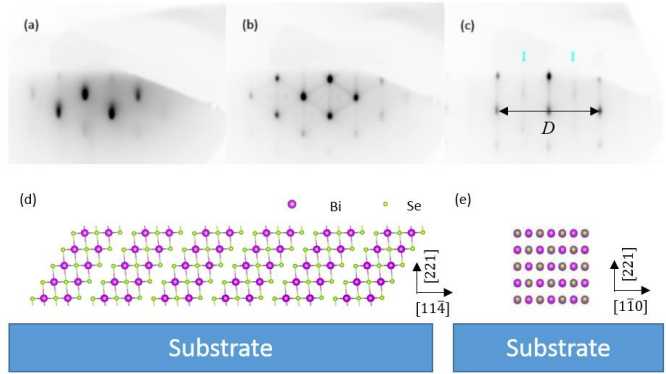

FIG. 1. RHEED patterns of (a) roughened GaAs(001), (b) In$_2$Se$_3$ buffer, and (c) 22 nm Bi$_2$Se$_3$ grown on top. (d,e) Schematic drawing of the lattice of Bi$_2$Se$_3$(221) viewed along [1$\bar{1}$0] and [11$\bar{4}$], respectively.

By symmetry analysis of the diffraction spots as well as by post-growth STM and TEM measurements of the sample, we establish that the epifilm is of high-index Bi$_2$Se$_3$(221). Fig. 2(a) shows a STM topographic image of the sample, which reveals a characteristic stripe morphology being distinctly different from that of Bi$_2$Se$_3$(111) surface. It resembles that of Bi$_2$Se$_3$(221) grown on InP(001) as reported earlier [20]. Such a striped morphology of Bi$_2$Se$_3$(221) surface is understood by the very different atomic bonding characteristics in the two orthogonal directions on surface: covalent along [1$\bar{1}$0] but containing vdW planes along [11$\bar{4}$] (refer to Fig. 1(d)). As a result, there is a huge anisotropy of surface growth rates between [1$\bar{1}$0] and [11$\bar{4}$] directions, giving rise to the stripe morphology as seen in Fig. 2(a).

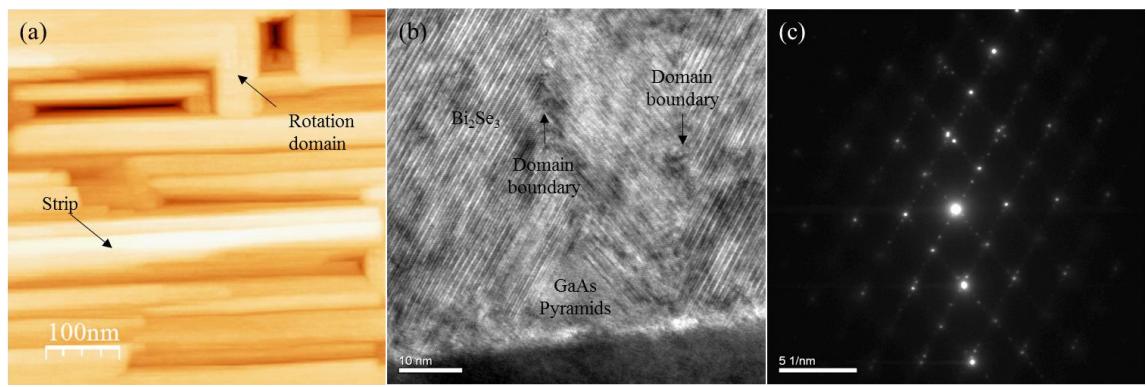

FIG. 2. (a) STM image of epitaxial Bi$_2$Se$_3$(221). Rotation domain and strip morphology are indicated by black arrows. (b) Cross-sectional TEM images of Bi$_2$Se$_3$(221) epifilm close to the interface taken along [1$\bar{1}$0]. GaAs pyramids are formed by the etching process. Two domain boundaries are marked by the black arrows. (c) SAED of the epifilm taken along the same direction as that of (b).

Fig. 2(b) presents the cross-sectional HRTEM image taken along [1$\bar{1}$0] zone axis of Bi$_2$Se$_3$(221) and the corresponding SAED image is given in Fig. 2(c). Again, the latter is consistent with the R$\bar{3}$m symmetry of a rhombohedral crystal with the vertical (surface normal) being [221]. The lattice images of Fig. 2(b) clearly resolves Bi$_2$Se$_3$ QLs separated by the vdW gaps, which are inclined with respect to interface normal. The angle of incline is close to that between (111) and (221) planes of Bi$_2$Se$_3$ (57.9°). The rough and {111} facets of the substrate and In$_2$Se$_3$ buffer are discernable, revealing the epitaxial relation of $(111)_{Bi_2Se_3}\|(111)_{In_2Se_3}\|(111)_{GaAs}$. Therefore, the high-index Bi$_2$Se$_3$(221) film grown on facetted GaAs(001) substrate is promoted and guided by the {111} plane of Bi$_2$Se$_3$ parallel to the {111} facets of the substrate. The presence of four equivalent {111} facet planes on GaAs(001) implies rotation domains to be unavoidable in the top Bi$_2$Se$_3$(221) epilayer, which can indeed be inferable from both STM and TEM measurements. However, anisotropy of growth or etching rates of (111) versus (1$\bar{1}$1) faces might lead to different sizes of the two facets and so a continuous growth of Bi$_2$Se$_3$ ultimately causes one domain to dominate over the other. The morphology thus becomes elongated stripes in one direction while the 90°-rotated stripes are of a smaller proportion as is evident from Fig. 2(a).

Bi$_2$Se$_3$(221) surface has dangling bonds pointing vertically out of the plane. At the hetero-interface between the deposit and the substrate, the interface interaction is covalent, which offers the possibility of straining the film by the lattice misfit between the epifilm and substrate. As mentioned earlier, for layered compounds like Bi$_2$Se$_3$, strain can be accommodated readily along [11$\bar{4}$] where it contains weakly couple vdW planes. Along the orthogonal [1$\bar{1}$0] direction, although the chemical bonds are covalent, high strains may still be tolerable due to the thin slab structure of such layered materials. Thus strain relaxation in this direction is expectedly slower even than that of a covalent crystal. We follow the evolution of in-plane lattice parameter $a_{[1\bar{1}0]}$ of epitaxial Bi$_2$Se$_3$(221) by the RHEED through real-time measurements of inter-diffraction streak spacing $D$. Specifically, the spacing $D$ between (01) and (0$\bar{1}$) diffraction streaks viewed along [11$\bar{4}$] direction of Bi$_2$Se$_3$ are extracted during film deposition, which are then translated into the lattice parameter $a$ of Bi$_2$Se$_3$ in [1$\bar{1}$0] direction according to $a \propto 1/D$.

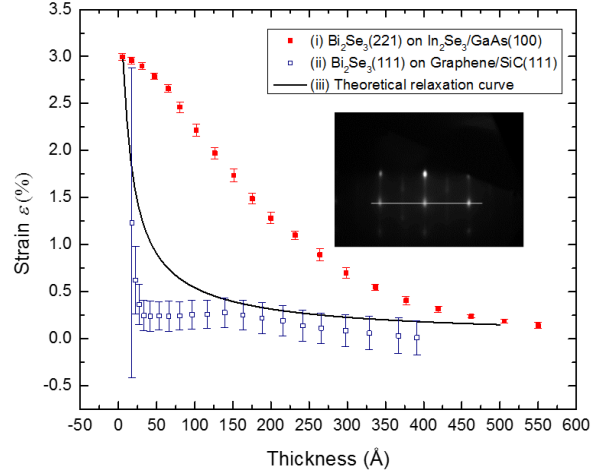

FIG. 3. Strain profile measured by the RHEED (refer to the inset) of epitaxial $Bi_2Se_3$(221) on $In_2Se_3$-buffered GaAs(001) (red solid symbols) and of $Bi_2Se_3$(111) grown on graphene (blue open symbols). The solid line is the theoretical relaxation curve of strain for a covalent epitaxial system.[26]

At room-temperature, the in-plane lattice misfit ($f$) between $Bi_2Se_3$ epifilm and $In_2Se_3$ buffer is $f = \frac{a_{Bi_2Se_3} - a_{In_2Se_3}}{a_{Bi_2Se_3}} \approx 3\%$. By elasticity theory, this lattice misfit strain may only be sustained in the film up to a critical thickness of $h_c \sim 0.15$ nm when uniform stress in two-dimensional interface plane is assumed (as for epitaxial growth of a covalent crystal).[26] Fig. 3 shows an experimental strain-relaxation profile of $Bi_2Se_3$(221) (curve i) as derived by $\varepsilon = \frac{a_0 - a}{a_0} = \frac{D - D_0}{D}$, where $a$ and $a_0$ are in-plane lattice constants, and $D$, $D_0$ are inter-diffraction streak spacing of the deposit ($a$ and $D$) and strain-free $Bi_2Se_3$ ($a_0$ and $D_0$), respectively. The measurements of $D$ is made along the white horizontal line drawn in the inset of Fig. 3. Although strain has indeed been seen to start to relax from very early stage of deposition, its relaxation profile is seen to be much slower than the prediction by elasticity theory for covalent crystals[26] (the solid line labelled as (iii) in Fig. 3). As noted earlier, in this direction, there is no strain-accommodating vdW plane, so one does not expect a fully strained film to persist with increasing thickness. However, because of the inclined QL-slab nature of the epifilm [cf. Fig. 1(d)], a higher strain can indeed be expected or its relaxation takes a slower rate.[26] This contrasts greatly the case of $Bi_2Se_3$(111) films grown on vdW substrate, such as graphene (e.g., curve (ii) in Fig. 3).[22] In particularly, strain relaxation in $Bi_2Se_3$(111) is much quicker than the theoretical prediction for growth of covalent crystal (compared (ii) and (iii) in Fig. 3), signifying the inability of strain-bearing at the vdW hetero-interfaces. This experiment, however, demonstrates that strained $Bi_2Se_3$ films can be achieved when grown along the off $c$-axis direction.

In order to characterize the strength of chemical bonding at the hetero-interface, we have performed first-principle calculations to examine the charge transfer at the interface. Firstly, the pure $Bi_2Se_3$(221) slab model was calculated and the resulted band structure reproduced the result of Ref.[18], revealing the anisotropic Dirac cone. Importantly, our calculation suggested that only the *p*-orbitals of Bi and Se atoms on surface contributed to the Dirac state.

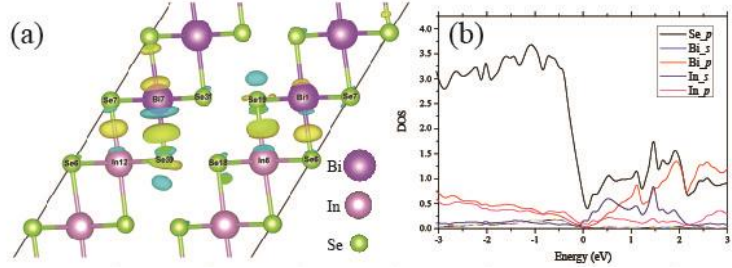

FIG. 4. Charge transfer and chemical bonding at $Bi_2Se_3$-$In_2Se_3$ (221) interface: (a) Ball-and-stick model of $Bi_2Se_3$-$In_2Se_3$ heterostructure along [221] superimposed with the differential charge distribution of the interface (isosurface level: 0.003 e/Bohr$^3$). The yellow region is where the charge density increases, while the light blue region is the opposite. (b) Total DOS as contributed from the *s*- and *p*-orbitals of all Bi, In, Se atoms as labelled in (a).

The hetero-interface between $Bi_2Se_3$ and $In_2Se_3$ along [221] is modeled as in Fig. 4(a). As the Bi, In and Se atoms at the heteointerface, which are the 'edge atoms' of truncated QLs, host unsaturated dangling bonds, chemical bonding of such atoms at the $Bi_2Se_3$/$In_2Se_3$ interface is expected, which manifests by a charge redistribution as revealed by the calculation result [Fig. 4(a)]. Specifically, a charge increase is seen midway between the atomic pairs like Bi-Se or In-Se, reflecting the covalent bonding of these atoms. Such chemical bonding at $Bi_2Se_3$/$In_2Se_3$ (221) interface represents a very different property from that of the vdW interface along [111]. From the derived DOS from different orbitals of atoms at the heterointerface [Fig. 4(b)], we may assert that the chemical bondings are mainly contributed from the *p*-orbitals of Bi ($Bi_p$), Se ($Se_p$) and the *s*-orbital of In ($In_s$). By projecting a set of DOS to each individual atom as labelled in Fig. 4(a), we further note that the $Bi_p$($Se_p$) orbitals in $Bi_2Se_3$ and the $Se_p$($In_s$) orbitals from $In_2Se_3$ form chemical bonds. Recalling that the Dirac state is contributed entirely from $Bi_p$ and $Se_p$ orbitals, it will be interesting to note if such chemical bonding may lead to modification and/or new effects of the Dirac states.[27,28]

To conclude, we have achieved a high-index in $Bi_2Se_3$(221) film by MBE on a faceted GaAs(001) substrate with an $In_2Se_3$ buffer. We note a retarded strain relaxation process in such a film along $[1\bar{1}0]$, which can be attributed to the inclined QL-slab configuration of the crystal. Such a strain (~ 3%) accessible by the experiment would bring some anisotropic tuning effects on Dirac electrons. At last, we have performed first-principle calculation to study the interaction between high-index $Bi_2Se_3$ and $In_2Se_3$, and a strong chemical bonding is revealed by the dramatic charge redistribution at the interface. We remark that such chemically bonded interface may bring some new effects which are not observed for vdW interface. This work thus has shed a new light on future strain related studies.

The work described in this paper was supported in full from a grant of the SRFDP and RGC ERG Joint Research Scheme of Hong Kong RGC and the Ministry of Education of China (No. M-HKU709/12). This research is conducted in part using the research computing facilities and/or advisory services offered by Information Technology Services, the University of Hong Kong.